\input psfig.sty

\def\ptitle{Smooth spectral transition from Coulomb to oscillator}
\nopagenumbers
\hsize 6.0 true in 
\hoffset 0.25 true in 
\emergencystretch=0.6 in                 
\vfuzz 0.4 in                            
\hfuzz  0.4 in                           
\vglue 0.1true in
\mathsurround=2pt                        
\topskip=20pt                            
\def\nl{\noindent}                       
\def\np{\hfil\vfil\break}                
\def\title#1{\bigskip\noindent\bf #1 ~ \trr\smallskip} 
\font\trr=cmr12                         
\font\bf=cmbx12                         
\font\sl=cmsl12                         
\font\it=cmti12                         
\font\trbig=cmbx12 scaled 1500          
\font\tiny=cmr10                         
\def\ma#1{\hbox{\vrule #1}}             
\def\mb#1{\hbox{\bf#1}}                 
\def\ng{>\kern -9pt|\kern 9pt}          
\def\bra{{\rm <}}                       
\def\ket{{\rm >}}                       
\def\hi#1#2{$#1$\kern -2pt-#2}          
\def\hy#1#2{#1-\kern -2pt$#2$}          

\def\sgn{{\rm sgn}}
\def\half{{1 \over 2}}


\output={\shipout\vbox{\makeheadline
                                      \ifnum\the\pageno>1 {\hrule}  \fi 
                                      {\pagebody}   
                                      \makefootline}
                   \advancepageno}

\headline{\noindent {\ifnum\the\pageno>1 
                                   {\tiny \ptitle\hfil page~\the\pageno}\fi}}
\footline{}
\newcount\zz  \zz=0  
\newcount\q   
\newcount\qq    \qq=0  

\def\pref #1#2#3#4#5{\frenchspacing \global \advance \q by 1     
    \edef#1{\the\q}
       {\ifnum \zz=1 { %
         \item{[\the\q]} 
         {#2} {\bf #3},{ #4.}{~#5}\medskip} \fi}}

\def\bref #1#2#3#4#5{\frenchspacing \global \advance \q by 1     
    \edef#1{\the\q}
    {\ifnum \zz=1 { %
       \item{[\the\q]} 
       {#2}, {\it #3} {(#4).}{~#5}\medskip} \fi}}

\def\gref #1#2{\frenchspacing \global \advance \q by 1  
    \edef#1{\the\q}
    {\ifnum \zz=1 { %
       \item{[\the\q]} 
       {#2}\medskip} \fi}}

 \def\sref #1{~[#1]}

\def\references#1{\zz=#1
   \parskip=2pt plus 1pt   
   {\ifnum \zz=1 {\noindent \bf References \medskip} \fi} \q=\qq
\pref{\quigg}{C. Quigg and J. L. Rosner, Phys. Lett.}{B 71}{153, 1977}{}
\pref{\mach}{M. Machacek and Y. Tomozawa, Ann. Phys. (N.Y.)}{110}{40, 1978}{}
\pref{\hallka}{R. L. Hall and C. S. Kalman, Phys. Lett.}{B 83}{80, 1979}{}
\pref{\marta}{A. Martin, Phys. Lett.}{B 100}{511, 1981}{}
\bref{\luch}{W. Lucha and F. F. Sch\"oberl}{Die Starke Wechselwirkung: Eine Einf\"uhrung in Nichtrelativistische Potentialmodelle}{Wissenschaftsverlag, Wien, 1989}{The Coulomb plus linear potential is discussed on p. 88.}
\pref{\martb}{A. Martin and J-M Richard, Phys. Lett.}{B 355}{345, 1995}{}
\pref{\god}{S. Godfrey and J. Napolitano, Rev. Mod. Phys.}{71}{1411 (1999)}{The Coulomb plus linear potential is discussed on p. 1418.}
\pref{\quigga}{C. Quigg and J. Rosner, Phys. Rep }{C56}{167, 1979}{}
\pref{\gas}{S. Gasiorowicz and J. Rosner, Am. J. Phys.}{49}{954, 1981}{}
\pref{\hallc}{R. L. Hall, J. Math. Phys.}{24}{324, 1983}{} 
\pref{\halld}{R. L. Hall, J. Math. Phys.}{25}{2078 (1984)}{} 
\pref{\hallsa}{R. L. Hall and N. Saad, J. Chem. Phys.}{109}{2983 (1998)}{}
\pref{\hallb}{R. L. Hall, Phys. Rev.}{A 39}{5500, 1989}{} 
\bref{\gel}{I. M. Gelfand and S. V. Fomin}{Calculus of Variations}{Prentice-Hall, Englewood Cliffs, 1963}{Legendre transformations are discussed on p. 72.} 
\pref{\hallf}{R. L. Hall, Phys. Lett. A}{265}{28 (2000)}{}

\pref{\halla}{R. L. Hall, J. Math. Phys.}{34}{2779 (1993)}{} 

\bref{\flug}{S. Fl\"ugge}{Practical Quantum Mechanics}{Springer, New York, 1974}{The linear potential is discussed on p. 101.}
\pref{\hallg}{R. L. Hall, J. Math. Phys.}{33}{1710 (1992)}{} 
\pref{\weyl}{Math. Ann. }{71}{441 (1911)}{}
\pref{\fan}{Ky Fan, Proc. Nat. Acad. Sci. (U.S.)}{35}{652 (1949)}{}
\bref{\wein}{A. Weinstein and B. Stenger}{Methods of Intermediate Problems for Eigenvalues}{Academic, New York, 1972}{Weyls' theorem is discussed on p. 163.}
\pref{\thira}{W. Thirring, Found. Phys.}{20}{1103, (1990)}{}
 \bref{\thirb}{W. Thirring}{A Course in Mathematical Physics 4: Quantum Mechanics of Large Systems}{Springer, New York, 1983}{}
\pref{\hallh}{R. L. Hall, Phys. Rev. A}{45}{7682 (1992)}{}
\pref{\halli}{R. L. Hall, Phys. Rev. A}{51}{3499 (1995)}{}
\pref{\halln}{R. L. Hall, Phys. Rev. D}{30}{433 (1984)}{}
\pref{\hallj}{R. L. Hall and H. R. Post, Proc. Phys. Soc (Lond.)}{90}{381 (1967)}{}
\pref{\hallk}{R. L. Hall, Proc. Phys. Soc (Lond.)}{91}{16 (1967)}{}
\bref{\prug}{E. Prugovecki}{Quantum Mechanics in Hilbert Space}{Academic, New York, 1981}{}
\bref{\reed}{M. Reed and B. Simon}{Methods of Modern Mathematical Physics IV: Analysis of Operators}{Academic, New York, 1978}{}
\bref{\thirc }{W. Thirring}{A Course in Mathematical Physics 3: Quantum Mechanics of Atoms and Molecules}{Springer, New York, 1981}{}
\pref{\halll}{R. L. Hall, Can. J. Phys.}{50}{305 (1972)}{}
\pref{\hallm}{R. L. Hall, Aequ. Math.}{8}{281 (1972)}{}
\pref{\hallo}{R. L. Hall, Phys. Rev. D}{37}{540 (1988)}{}
\pref{\kang}{J. S. Kang and H. J. Schnitzer, Phys. Rev. D}{12}{841 (1975)}{}
\pref{\grom}{D. Gromes and I. O. Stamatescu, Nucl. Phys. B}{122}{213 (1976)}{}

 }

\references{0}    

\trr 
\vskip 1.0true in
\centerline{\trbig Smooth spectral transition from}
\vskip 0.5true in
\centerline{\trbig Coulomb to oscillator}
\vskip 0.5true in
\baselineskip 12 true pt 
\centerline{\bf Richard L. Hall}\medskip
\centerline{\sl Department of Mathematics and Statistics,}
\centerline{\sl Concordia University,}
\centerline{\sl 1455 de Maisonneuve Boulevard West,}
\centerline{\sl Montr\'eal, Qu\'ebec, Canada H3G 1M8.}
\vskip 0.2 true in
\centerline{email:\sl~~rhall@cicma.concordia.ca}
\bigskip\bigskip

\baselineskip = 18true pt  
\centerline{\bf Abstract}\medskip
Non-relativistic potential models are considered of the pure power $V(r) = \sgn(q) r^{q}$ and logarithmic $V(r) = \ln(r)$ types.  Envelope representations and kinetic potentials are employed to show that these potentials are actually in a single family. The log spectra can be obtained from the power spectra by the limit $q\rightarrow 0$ taken in a smooth representation $P_{n\ell}(q)$ for the eigenvalues $E_{n\ell}(q).$  A simple approximation formula is developed which yields the first thirty eigenvalues with error $< 0.04\%.$  Extensions to potentials with linear combinations of terms such as $-a/r + br$ and applications to spatially-symmetric few-body problems are discussed. 
\medskip\noindent PACS~~12.39.Pn;~12.39.Jh;~03.65.Ge.

\np
  \title{1.~~Introduction}
We consider first a single particle that moves in a central potential $V(r)$ and obeys non-relativistic quantum mechanics.  We study two cases: (a) $V(r) = \sgn(q)r^{q},$ and (b) $V(r) = \ln(r).$  We shall show that these two problems are intimately related and that this allows us to construct a single formula which yields accurate approximations for the corresponding discrete Schr\"odinger eigenvalues.  The power-law and logarithmic potentials continue to be employed as non-relativistic models for quark confinement\sref{\quigg-\god}. Hence it is important to understand that they are, in a sense, from the same family of potentials. A detailed analysis of the properties of non-relativistic potential models is given in the review article by Quigg and Rosner\sref{\quigga}. The approach to this question based on the expression $\{dr^{q}/dq\}_{q\rightarrow 0} = \ln r$ is also been discussed in Ref.[\gas]. In the present paper we approach the same problem by using what we call the `$P$-representation' for the discrete spectrum generated by power-law potentials. This allows us to construct a general energy eigenvalue formula which joins $q = -1$ smoothly to $q = 2,$ passing through the logarithmic case $q = 0.$\medskip

The power-law family and logarithmic potential will be kept distinct until it becomes clear in what sense the logarithmic potential corresponds to the limit $q \rightarrow 0$ of the power-law class.  We first treat the elementary issue of scaling.  We write the eigenvalues of the bare problems as:

$$\cases{
-\Delta + \sgn (q)r^{q} & $\longrightarrow E_{n\ell}(q)$\cr
&\cr
-\Delta + \ln (r)       & $\longrightarrow E_{n\ell}^{L},$\cr
}\eqno{(1.1)}$$

\nl where $n = 1,2,3,\dots$ counts the discrete eigenvalues in each angular-momentum subspace labelled by $\ell = 0,1,2,\dots.$ The eigenvalues so labelled have degeneracy precisely $2\ell+1.$  Scaling arguments\sref{\hallc,\hallka} can be used to show that more general Hamiltonians have the corresponding eigenvalues given in terms of $\{E_{n\ell}(q),~E_{n\ell}^{L}\}$ by

$$\cases{
-\omega\Delta + v\ \sgn (q)r^{q} & $\longrightarrow \omega\left ({v\over\omega}\right)^{2\over{2+q}}E_{n\ell}(q)$\cr
&\cr
-\omega\Delta + v\ln (r)       & $\longrightarrow vE_{n\ell}^{L} - \half v\ln\left({v\over\omega}\right),$\cr
}\eqno{(1.2)}$$

\nl where $v > 0$ is a coupling parameter and $\omega > 0$ might, for example, be $\omega = \hbar^{2}/(2m).$  As a consequence of these scaling rules we shall only need to consider the special case $\omega = v = 1$ for this family of potentials.\medskip

In Section~(2) we review the concepts of envelope representations and kinetic potentials\sref{\hallc, \halld}.  These geometrical methods allow us to construct an exact semi-classical representation for the discrete spectra of Schr\"odinger operators in which the eigenvalues are recovered by a minimization over a single real kinetic-energy variable.  In Section (3) we show, for the power-law family of potentials, that this leads, in particular, to the smooth representation $P_{n\ell}(q)$ for the eigenvalues $E(q)$ (with $v = 1$) which, in turn, allows us to show that the log spectrum can be derived as a limit from the family of power spectra.  As a motivation for this, we plot the power eigenvalues $E_{n\ell}(q)$ as functions of $q$ in Fig.(1).  These graphs are to be compared with the corresponding smooth $P$-representation shown in Fig.(2).  Once the $P$-representation is established, we go on to consider the {\it log-power theorem} $P_{n\ell}(0) = P_{n\ell}^{L},$ and, in Section~(4), to develop an approximation formula for $P_{n\ell}(q).$ In Section (5) we consider potentials which are linear combinations of powers, such as the Coulomb plus linear potential $-a/r + br;$ in Section (6) we use these results to study few-body problems for which spatially-symmetric states are accessible.

  \title{2.~~Envelope representations and kinetic potentials}
We distinguish a potential $V(r) = vf(r)$ from its shape $f(r),$ where the positive parameter $v$ is often called the `coupling constant'.  The idea behind envelope representations is suggested by the question: if one potential (shape) $f(r)$ can be written as a smooth transformation $f(r) = g(h(r))$ of another $h(r),$  what spectral relationship might this induce?  We consider potential shapes that support at least one discrete eigenvalue for sufficiently large values of the coupling $v$ and suppose for definiteness that the lowest eigenvalue of $-\Delta + vh(r)$ is given by $H(v)$ and that of $-\Delta + vf(r)$ by $F(v).$  If the transformation function $g(h)$ is smooth, then each tangent to $g$ is an affine transformation of the `envelope basis' $h$ of the form $f^{t}(r) = \alpha(t)h(r) + \beta(t),$ where $r = t$ is the point of contact. The coefficients $\alpha(t)$ and $\beta(t)$ are obtained by demanding that the `tangential potential' $f^{(t)}(r)$ and its derivative agree with $f(r)$ at the point of contact $r = t.$  Thus we have
$$\alpha(t) = f'(t)/h'(t),\quad \beta(t) = f(t) - \alpha(t)h(t).\eqno{(2.1)}$$
\nl The geometrical configuration is illustrated in Fig.(3) in which $f(r) = -1/r + r$ and the envelope basis for the upper family is the linear potential $h(r) = r$ and, for the lower family, the Coulomb potential $h(r) = -1/r.$  The spectral function for the tangential potential $f^{(t)}(r) = \alpha(t)h(r) + \beta(t)$  is given by $H^{(t)}(v) = H(v\alpha(t)) + v\beta(t).$ If the transformation $g(h)$ has definite convexity, say $g''(h) > 0,$ then each tangential potential lies beneath $f(r)$ and, as a consequence of the comparison theorem, we know that each corresponding tangential spectral function $H^{(t)}(v),$ and the envelope of this set, lie beneath $F(v).$   Similarly, in the case where $g$ is concave ($g''(h) < 0$), we obtain upper bounds to $F(v).$  These purely geometrical arguments, depending on the spectral comparison theorem, extend easily to the excited states.  The spectral curves corresponding to the envelope representations for the potential in Fig.(3) are shown in Fig.(4) for the excited state $(n, \ell) = (2, 4).$  For comparison the exact curve $E = F(v)$ is also shown; this curve will be close to the Coulomb envelope for large $v$ and to the linear envelope for small $v.$  Of course, the envelopes of which we speak still have to be determined explicitly.

Whilst extensions of this idea to completely new problems, such as simultaneous transformations of each of a number of potential terms\sref{\hallsa}, are best formulated initially with the basic argument outlined above, the question arises, in the \hi{1}{term} case, as to whether there is a simple way of determining the envelopes of the families of upper and lower spectral functions.  One solution of this problem is by the use of `kinetic potentials' which were introduced\sref{\hallc, \halld} precisely for this purpose.  The idea is as follows.  To each spectral function $F_{n \ell}(v)$ there is a corresponding `kinetic potential' (= {\it minimum mean iso-kinetic potential}) $\bar{f}_{n \ell}(s).$ The relationship between $F_{n \ell}$ and $\bar{f}_{n \ell}$ is invertible and is essentially that of a Legendre transformation\sref{\gel}: we can prove in general that $F$ is concave, $\bar{f}(s)$ is convex, and 
$$F''(v)\bar{f}''(s) = -{1 \over {v^{3}}}.\eqno{(2.2)}$$
\nl The explicit transformation formulas are as follows 
$$\{\bar{f}_{n \ell}(s) = F'_{n \ell}(v),\quad s = F_{n \ell}(v)-vF'_{n \ell}(v)\}\quad\leftrightarrow\quad
\{F_{n \ell}(v)/v = \bar{f}_{n \ell}(s) - s\bar{f}'_{n \ell}(s),\quad 1/v = - \bar{f}'_{n \ell}(s)\}.\eqno{(2.3)}$$
\nl An {\it a priori} definition of the ground-state kinetic potential $\bar{f}_{1 0}(s) = \bar{f}(s)$ is given by 
$$\bar{f}(s) = \inf_{{{\scriptstyle \psi \in {\cal D}(H)} \atop {\scriptstyle (\psi,\psi) = 1}} \atop {\scriptstyle (\psi, -\Delta\psi) = s}} (\psi, f\psi),\eqno{(2.4)}$$
\nl where ${\cal D} \subset L^{2}(\ma{R}^{3})$ is the domain of the Hamiltonian.  The definition for the excited states is a little more complicated\sref{\halld} and, in view of (2.3), will not be needed in what follows.\smallskip

What is important is that the spectral functions, either exact or approximate, are recovered from the corresponding kinetic potentials by a minimization over the kinetic-energy variable $s.$  The total minimization required by min-max has been divided into two steps: the first is constrained by $(\psi,-\Delta\psi) = s$ and the second is a minimization over $s.$  We have in all cases:
$$F_{n \ell}(v) = \min_{s > 0}\left\{s + v\bar{f}_{n \ell}(s)\right\}.\eqno{(2.5)}$$
\nl Another form of this expression is possible for the kinetic potential is monotone and allows us to change variables $(s\rightarrow r)$ by $f(r) = \bar{f}_{n \ell}(s).$  Thus we have
$$F_{n \ell}(v) = \min_{r > 0}\left\{K^{(f)}_{n \ell}(r) + vf(r)\right\},\quad K^{(f)}_{n \ell} = \bar{f}_{n \ell}^{-1}\circ f.\eqno{(2.6)}$$
\nl The two corresponding expressions of the envelope approximation then become
$$f_{n \ell}(s) \approx f(h_{n \ell}(s)),\quad {\rm or}\quad K^{(f)}_{n \ell}\approx K^{(h)}_{n \ell}.\eqno{(2.7)}$$
\nl The second form (2.6) of the expression for $F_{n \ell}(v)$ isolates the potential shape $f$ itself and leads to an inversion sequence\sref{\hallf} which reconstructs the potential from a single given spectral function; but this is another story.

It is useful here to provide the formulas for the kinetic potentials corresponding to pure power-law potentials.  Since, by elementary scaling arguments, we have
$$-\Delta + v\ {\rm sgn}(q)r^q\quad \rightarrow \quad F(v) = F(1)v^{2\over{q+2}},\eqno{(2.8)}$$
\nl where $F(1) = E(q)$ is the bottom of the spectrum of $-\Delta + {\rm sgn}(q)r^q,$ we immediately find from (2.3) that the kinetic potentials for these potentials are given by
$$\bar{f}_{n\ell}(s) = \left({2\over {q}}\right)\left|{{qE_{n\ell}(q)}\over{2+q}}\right|^{{q+2}\over 2}\ s^{-{q\over{2}}}.\eqno{(2.9)}$$
\nl Meanwhile the corresponding \hi{K}{functions} all have the same simple form
$$K_{n\ell}^{(q)}(r) = \left({{P_{n\ell}(q)}\over {r}}\right)^{2},\eqno{(2.10)}$$
\nl where the $P$ numbers are given by 
$$P_{n\ell}(q) = \left|E_{n\ell}(q)\right|^{{2+q}\over{2q}}\left[{{2}\over{2+q}}\right]^{{1}\over{q}}\left|{{q}\over{2+q}}\right]^{\half},\quad q\neq 0.\eqno{(2.11)}$$
\nl Consequently the power-law kinetic potentials (2.9) may be expressed in the simple form 
$$\bar{f}_{n\ell}(s) = {\rm sgn(q)}\left({{P_{n\ell}(q)}\over {s^{\half}}}\right)^{q},\quad q\neq 0.\eqno{(2.12)}$$
\nl Meanwhile, as we shall see, for the log potential $f(r) = \log(r)$ we have $\bar{f}_{n\ell}(s) = \log(P_{n\ell}(0)/s^{\half}).$ These formulas are particularly useful for the analysis of {\it sums} of powers discussed in Section (5) below.

\np 
  \title{3.~~The functions $E(q)$ and $P(q)$ for pure powers and the $\log$ potential}
By using the $P$-representation for the spectrum we have from (2.6) and (2.10)
$$E_{n\ell} = \min_{r > 0}\left\{{{P_{n\ell}^2(q)}\over {r^2}} + V(r)\right\}.\eqno{(3.1)}$$
Independently of the theory discussed in Section (2), we see that this representation is well defined provided we can always construct the $P$ which corresponds to a given $E.$  For smooth potentials, the minimum exists and is unique if $r^{3}V'(r)$ is monotone, a condition which is certainly satisfied if $V(r)$ is a pure power or log.  For these potentials, Eq.(3.1) is therefore an exact semi-classical representation for the eigenvalues: the kinetic energy term scales like $L^{-2},$ as it should. It is perhaps interesting also to note that positive factors such as $\omega$ and $v$ can be re-introduced in front of the kinetic- and potential-energy terms, without having to revise the $P$ numbers.  For the specific pure-power and log problems at hand, we have respectively:
$$\eqalign{E_{n\ell}(q) &= \min_{r > 0}\left\{{{P_{n\ell}^2}(q)\over {r^2}} + \sgn(q)r^{q}\right\}\cr
&= \sgn(q)\left({q\over 2} + 1\right)\left({{2P_{n\ell}^{2}(q)}\over {|q|}}\right)^{q\over {q+2}}\cr}\eqno{(3.2)}$$
and
$$\eqalign{E_{n\ell}^{L} &= \min_{r > 0}\left\{{{P_{n\ell}^L}^{2}\over {r^2}} + \ln(r)\right\}\cr
&= \ln\left((2e)^{\half} P_{n\ell}^{L}\right).\cr}\eqno{(3.3)}$$

\nl Some of the $E,$ and the corresponding $P,$ are already known exactly, from elementary quantum mechanics. Thus from the known eigenvalues for the Hydrogen atom $E_{n\ell}(-1) = -[2(n+\ell)]^{-2}$ and the harmonic oscillator $E_{n\ell}(2) = 4n + 2\ell -1$ we immediately obtain the outer $P$ numbers of our range of interest $-1 \leq q \leq 2:$

$$P_{n\ell}(-1) = n+\ell,\quad P_{n\ell}(2) = 2n + \ell - \half.\eqno{(3.4)}$$

\nl Our goal is now to interpolate $P_{n\ell}(q)$ at interior points of the range.  First we use kinetic potentials to establish two theoretical results.  It is straightforward to show that each power potential with power $q_1$ is a convex transformation of all other power potentials with powers $q_2 < q_1.$  Meanwhile, the log potential is a convex function of each negative power, and, at the same time, it is a concave transformation of each positive power.  Two results which follow from this analysis are (i)\sref{\hallb} the functions $P_{n\ell}(q)$ are monotone increasing in $q,$  and (ii)\sref{\halla} the {\it log-power theorem}, which says that
$\lim_{q \rightarrow 0}P_{n\ell}(q) = P_{n\ell}(0) = P_{n\ell}^{L}.$ 

As is clear from Fig.(1), the spectra $E_{n \ell}(q)$ of the power potentials converge on the values $\pm 1$ as the power $q$ approaches zero respectively from positive or negative values. Meanwhile, the set of points for the $\log$ spectrum $E_{n\ell}^{L}$ would appear to be quite unrelated, and would have no natural place on this graph.  In the $P$-representation the picture is quite different, as is shown in Fig.(2).  The monotone $P(q)$ curves are distinct at $q = 0,$ and we know that they are exactly equal there to the values corresponding to the log potential.  This smooth expression of the entire spectral family suggests that we can interpolate the $P$ curves with a simple polynomial, a task which is taken up in the next Section. We made an earlier attempt at an interpolation\sref{\hallb}, which was much less accurate because it could not benefit from the (yet undiscovered) log-power theorem. 
  \title{4.~~Interpolation formula for $P(q)$}
As a model for the $P$ curves we use a cubic polynomial which allows us to fit values at the four points $q = \{-1,0,1,2\};$ the exact formulas (3.4) are used for $q = \{-1, 2\}.$ We make the expansion about $q = 0$ so as to favour the $q$ values between log and linear that are used in potential models.  Thus we define
$$P(q) = a + b q + c q^{2} + d q^{3},\eqno{(4.1)}$$
\nl in which, for simplicity, we have omitted the quantum-number subscripts $n\ell.$  By inversion we have
$$\left[\matrix{a\cr b\cr c\cr d}\right] = \left[
\matrix{
0&1&0&0\cr
-{1\over 3} &-{1\over 2}&1&-{1\over 6}\cr
\half &-1&\half &0\cr
-{1\over 6}&\half&-\half&{1\over 6}
}\right]\left[\matrix{P(-1)\cr P(0)\cr P(1)\cr P(2)}\right]\eqno{(4.2)}.$$
\nl For the linear potential $q = 1$ the exact S-state eigenvalues may be expressed in terms of the zeros of the Airy function\sref{\flug}. We take these known values and complement them with others computed numerically for the log and linear potentials.  By inverting Eq.(3.2) and Eq.(3.3) we have thus computed the coefficients $P(0)$ for $\log$ and $P(1)$ for linear.   The $P$-data for the first 30 eigenvalues are exhibited in Table(1), along with the approximations we get for the energies $E_{n\ell}(\half).$   For comparison, we have tabulated also the percentage errors in these eigenvalue results.  For the first 30 eigenvalues, the percentage errors are all positive and less than $0.04\%$ at $q = \half;$  as $q$ approaches $q = 0$ or $q = 1,$ these errors decrease dramatically.  Similar accuracy is obtained for the whole range $-1 \leq q \leq 2$ of the interpolation.
\np 
  \title{5.~~Linear combinations}
The kinetic-potential formalism turns out to be helpful for another class of potential composition, namely linear combinations.
  Consider the bottom of the spectrum of the problem with hamiltonian
$$H = -\Delta + af^{(1)}(r) + bf^{(2)}(r).\eqno{(5.1)}$$
\nl Suppose that $\psi$ is the exact normalized ground-state wave function corresponding to eigenvalue $E$ and
 that $0<\omega<1,$ then we have exactly
$$ H = \omega\left(-\Delta + {a\over {\omega }}f^{(1)}(r)\right)
 + (1-\omega)\left(-\Delta + {b\over {1-\omega }}f^{(2)}(r)\right)$$
\nl and consequently for $E = (\psi,H\psi)$ we find the lower bound
$$E \geq \omega F^{(1)}({a\over {\omega }}) + (1-\omega) F^{(2)}({b\over {1-\omega }}).\eqno{(5.2)}$$
\nl By maximizing the right-hand side of (5.2) with respect to $\omega$ we obtain a best lower bound to $E$ expressed in terms of the `component' eigenvalues.  It turns out\sref{\halld} that this lower bound (after optimization over $\omega$) is given exactly by applying the kinetic-potential rule
$$\bar{f}(s) > a\bar{f}^{(1)}(s) + b\bar{f}^{(2)}(s).\eqno{(5.3)}$$
\nl This lower bound has been extended to arbitrary linear combinations and to continuous mixtures\sref{\hallg}; from the argument sketched above it is clear that the same bound is valid for the bottom of each angular-momentum subspace.  What is perhaps surprising is that for the other excited states it remains a `good' approximation (error a few \%).  It is in effect an approximate generalization of the classical operator sum theorem of Weyl\sref{\weyl - \wein}. The recipe we obtain for the eigenvalues of the Hamiltonian $H = -\Delta + \sum_{q}a(q){\rm sgn}(q)r^{q}$ for $a(q) > 0$ becomes, from (2.5) and (2.12), with the change of variable $s = 1/r^2,$ as follows
$$E_{n\ell} \approx \min_{r > 0}\left[{1\over{r^{2}}} + \sum_{q}a(q){\rm sgn(q)}\left(P_{n\ell}(q)r\right)^{q}\right].\eqno{(5.4)}$$
\nl This expression is an energy lower bound for the bottom of each angular-momentum subspace $n = 1.$  Thus kinetic potentials are almost additive; for the bottom of each angular-momentum subspace they are sub-additive; whenever the sum has only one term, the result is exact.
 
As an example we consider the Coulomb plus linear potential $f(r) = -a/r + br$ and find
$$E_{n\ell} \approx \min_{r > 0}\left[{1\over {r^2}} - {{a}\over{\nu r}} + b\mu r\right],\eqno{(5.5)}$$
\nl where we have a lower bound for Coulomb envelopes $\nu = \mu = P_{n\ell}(-1) = n+\ell,$ we have an upper bound with linear envelopes $\nu = \mu = P_{n\ell}(1),$ and we obtain an approximation by using the sum approximation $\nu = P_{n\ell}(-1),\quad \mu = P_{n\ell}(1),$  which latter is a lower bound whenever $n = 1.$  It is interesting that an upper bound to the ground-state energy $E_{1 0}$ derived from a scale optimized Gaussian wave function is  also provided by the same formula (5.5) if we assign
$$\nu = \left({{3\pi}\over{8}}\right)^{\half} = 1.085401,\quad \mu = \left({{6}\over{\pi}}\right)^{\half} = 1.381977.\eqno{(5.6)}$$
The Gaussian upper bound will be important for the \hi{N}{body} problem discussed in the next section.

Some progress can even be made towards a formula for these approximations\sref{\halln}.  By scaling arguments we have for every discrete eigenvalue ${\cal E}(\omega, a,b)$ of 
$$H = -\omega\Delta -{a\over r} + br,\quad\rightarrow\quad {\cal E}(\omega, a, b) = {{a^2}\over \omega}{\cal E}\left(1, 1, \lambda\right),\quad \lambda = {{b\omega^2}\over{a^{3}}}.\eqno{(5.7)}$$
By applying (5.5) with $a = 1$ and $b = \lambda$ we obtain an expression for $E(\lambda)$ whose inverse can be written explicitly as
$$\lambda = {{2(\nu E)^{3} -\nu E^{2}[(1+3\nu^{2}E)^{\half}-1]}\over{\mu[(1+3\nu^{2}E)^{\half} - 1]^{3}}},\eqno{(5.8)}$$
\nl where the values of $\nu$ and $\mu$ for the various bounds and approximations are given above, and $E\geq -1/4\nu^{2}.$  This is a global formula valid for all positive $a$ and $b.$
  \title{6.~~\hi{N}{body} problems}
We now discuss an application to \hi{N}{body} problems for which a spatially-symmetric state is accessible, such as a simple non-relativistic model for a \hi{3}{quark} system representing a nucleon.  Some general features and results for specific many-body systems may be found in Refs.[\thira, \thirb]. The Hamiltonian, with center-of-mass removed, for a system of $N$ identical particles each of mass $m$ interacting via central pair potentials may be written
$${\cal H} ={1 \over {2m}}\sum_{i = 1}^{N}\mb{p}_{i}^{2} -
{1 \over {2mN}}\left(\sum_{i = 1}^{N}\mb{p}_{i}\right)^{2} +
  \sum_{j>i=1}^{N}V_{o}f\left({{|\mb{r}_{i} -\mb{r}_{j}|} \over a}\right),\eqno{(6.1)}$$
where $V_{o}$ and $a$ are respectively the depth and range parameters of the potential with shape $f.$  By algebraic rearrangement (6.1) may be rewritten in the more symmetrical form 
$${\cal H} =\sum_{j>i=1}^{N}\left\{{1 \over {2mN}}(\mb{p}_{i} - \mb{p}_{j})^{2} + V_{o}f\left({{|\mb{r}_{i} -\mb{r}_{j}|} \over a}\right)\right\}.\eqno{(6.2)}$$
We now define new coordinates by $\rho = BR,$ where $\rho = [\rho_{i}]$ and $R = [\mb{r}_{i}]$  are column vectors of the new and old coordinates, respectively, and $B$ is a real constant $N\times N$ matrix.  For convenience we require all the rows of $B$ to be unit vectors, we let the elements of the first row all be equal to ${1 \over \sqrt{N}},$ so that $\rho_{1}$ is proportional to the centre-of-mass coordinate; we also require that the remaining $N-1$ rows of $B$ be orthogonal to the first row, so that they define a set of $N-1$ relative coordinates.  One more row is also fixed so that we have at least one pair distance at our disposal, namely
$$\rho_{2}= {{\mb{r}_{1} - \mb{r}_{2}} \over \sqrt{2}}.\eqno{(6.3)}$$
For spatially-symmetric states, we have found that Jacobi relative coordinates, for which $B$ is orthogonal, are the most useful.  Thus, corresponding to the transformation $\rho = BR$ of the coordinates, it follows that the column vector $P$ of the associated momenta transforms to the new momenta $\Pi = [\pi_{i}]$ by the relation $\Pi = (B^{T})^{-1}P = P.$  If $\Psi$ is any translation-invariant wave function for the \hi{N}{body} system composed of identical bosons, then we can write\sref{\hallj, \hallk} the following mean energy relation between the \hi{N}{body}  and \hi{2}{body} systems:
$$(\Psi, {\cal H}\Psi) = (\Psi, \ma{H} \Psi),\eqno{(6.4)}$$
where the `reduced'  two-particle Hamiltonian $\ma{H}$ is given by
$$\ma{H} = (N-1)\left({1 \over {2m}}\pi_{2}^{2} +
    {N \over 2}V_{o}f\left({{\sqrt{2}|\rho_{2}|} \over a}\right)\right).\eqno{(6.5)}$$

Further simplifications can be achieved if we work with dimensionless quantities.  We suppose that the translation-invariant \hi{N}{body} energy is ${\cal E}$ and we define the dimensionless energy and coupling parameters $E$ and $v$ by the expressions
$$E = {{m{\cal E}a^{2}} \over {(N-1)\hbar^{2}}},\quad
            v = {{NmV_{o}a^{2}} \over {2\hbar^{2}}}.\eqno{(6.6)}$$
It is then natural to define a dimensionless version of the reduced \hi{2}{body} Hamiltonian $\ma{H}$ and the relative coordinate $\rho_{2}$ by the relations
$$H ={{m\ \ma{H}a^{2}} \over {(N-1)\hbar^{2}}}= -\Delta + vf(r),\quad 
\mb{r} = \sqrt{2}\rho_{2}/a = \mb{r}_{1} - \mb{r}_{2},\quad r = \|\mb{r}\|.\eqno{(6.7)}$$
We note that the Hamiltonian $H$ depends on $N$ {\it only} through the dimensionless coupling parameter $v.$  By the Rayleigh-Ritz (min-max) principle\sref{\prug-\thirc}, we have the following characterization of the \hi{N}{body} ground-state energy parameter $E$ in terms of $H:$
$$E = \min_{\Psi}{{(\Psi, H\Psi)} \over {(\Psi, \Psi)}} = F_{N}(v),\eqno{(6.8)}$$
where $\Psi$ is a translation-invariant function of the $N-1$ relative coordinates (and spin variables, if any) which is symmetric or antisymmetric under the permutation of the $N$ individual-particle indices.  The \hi{N}{body} energy ${\cal E}$ is recovered from $E$ by inverting (6.6). Thus we have explicitly:
$${\cal E} = {{(N-1)\hbar^{2}} \over {ma^{2}}}F_{N}\left({{NmV_{o}a^{2}} \over {2\hbar^{2}}}\right).\eqno{(6.9)}$$

The energy bounds we use are summarized in terms of the $F$ functions
$$F_2(v) \leq E = F_N(v) \leq F_\infty(v) \leq F_G(v).\eqno{(6.10)}$$
The history of the equivalent \hi{2}{body} method for \hi{N}{particle} systems has been described in earlier papers\sref{\hallh-\hallj} and in the references therein.  The main result is a general energy lower bound which, with orthogonal Jacobi relative coordinates, is given by 
$$F_{2}(v)\leq E  = F_{N}(v),\eqno{(6.11)}$$
where $F_{2}(v)$ is the lowest eigenvalue of the \hi{1}{particle} (`reduced' \hi{2}{particle}) Hamiltonian $H = -\Delta + vf(r).$  With equal simplicity, our upper bound $F_{G}(N),$  provided (for spatially-symmetric states) by a  Gaussian trial function $\Psi,$ may also be expressed in terms of the \hi{1}{body} operator $H.$  This is a consequence of the following argument. If and only if the spatially-symmetric translation-invariant function $\Psi$ is Gaussian\sref{\halll, \hallm}, it may be factored in the form
$$\Psi(\rho_2,\rho_3,\dots,\rho_N) = \psi(\rho_2)\eta(\rho_3,\rho_4, \dots, \rho_N).\eqno{(6.12)}$$
But the equivalence (6.4) then implies, in this case, that $E \leq (\psi,H\psi)||\psi||^{-2}.$  This explains why the inequalities (6.10) collapse together in the case of the harmonic oscillator to the common value $E = 3v^{\half}$ for which the exact \hi{1}{body} lowest eigenfunction of $H$ is also Gaussian. In this argument we assume for the upper bound that $\bra H \ket$ has been optimized with respect to the scale of the wave function. The symmetry of these Gaussian functions is demonstrated most clearly by the following algebraic identity:
$$\sum_{j > i = 1}^{N}(\mb{r}_{i} - \mb{r}_{j})^{2} = N \sum_{k = 2}^{N}\rho_{k}^{2}.\eqno{(6.13)}$$ 

As an example\sref{\hallo} we consider a model symmetric \hi{N}{quark} system and we take some
 physical numbers from two early papers by Kang and Schnitzer\sref{\kang}, and Gromes 
and Stamatescu\sref{\grom}, and allow $2 \leq N \leq 5$ for this illustration, although $N > 3$
 no longer corresponds to a physical system.  The units are $\hbar = c = 1$ and we assume equal
 masses $m = 0.3 GeV$ and Coulomb-like coupling $a = 0.35.$  We keep $a$ fixed and plot the energy 
obtained from the formula (5.5) with $\nu = 1$ and $\mu = 2|E_{1 0}(1)/3|^{3/2} = 1.376083$ for
 a lower bound, and the same formula with the Gaussian values (5.6) for an upper bound.
  The energy curves as a function of the linear coupling $b$ are shown for each $N$ in Figs.(5)
 and (6), which only differ in the range of $b.$  One can only expect to get modest benefit from
 such a simple model.  However, definite energy bounds could provide a small island of security in an 
 otherwise highly complicated many-body environment.  

  \title{7.~~Conclusion}
In spite of present-day computing convenience, it is still very useful to have an approximate eigenvalue formula, particularly a simple and accurate one.  The possibility of a simple formula is a consequence of the existence of the smooth monotone $P$-representation for the eigenvalues.   Since the early work on potential models there has been a realization that, from a practical point of view, a log and a power potential, with $0 < q < 1,$ could serve in the model almost equally well, especially if $q$ is small.  The log-power theorem provides a theoretical basis for these concrete spectral observations. Once we have the smooth $P$ representation for the power-law and log eigenvalues we are able to provide spectral information concerning linear combinations of such terms and also for \hi{N}{body} problems such as non-relativistic models for systems of few quarks.  
   \title{Acknowledgment}
Partial financial support of this work under Grant No. GP3438 from the Natural Sciences and Engineering Research Council of Canada, and hospitality of the Erwin Schr\"odinger International Institute for Mathematical Physics in Vienna is gratefully acknowledged. 
\np
 \references{1}
\np
\noindent {\bf Table 1}~~The `input' values of $P_{n\ell}(0)$ and $P_{n\ell}(1);$ and the approximations $E^{A}_{n\ell}(\half)$ for $E_{n\ell}(\half)$ obtained via the cubic $P$ formula (3.1), with the percentage errors.
\baselineskip=16 true pt 
\def\vr{\vrule height 12 true pt depth 6 true pt}
\def\vra{\vr\hfill} \def\vrb{\hfill &\vra} \def\vrc{\hfill & \vr\cr\hrule}
\def\vrq{\vr\quad} 

$$\vbox{\offinterlineskip
 \hrule
\settabs
\+ \vrq \kern 0.4true in &\vrq \kern 0.4true in &\vrq \kern 0.7true in &\vrq \kern 0.7true in &\vrq \kern 0.7true in &\vrq \kern 0.7true in &\vrq \kern 0.5true in &\vr\cr\hrule
\+ \vra $n$ \vrb $\ell$\vrb $P_{n\ell}(0)$\vrb $P_{n\ell}(1)$\vrb $E^{A}_{n\ell}(\half)$\vrb $E_{n\ell}(\half)$\vrb \% \vrc
\+ \vra 1\vrb 0\vrb 1.21867\vrb 1.37608\vrb 1.83375\vrb 1.83339\vrb 0.019\vrc
\+ \vra 2\vrb 0\vrb 2.72065\vrb 3.18131\vrb 2.55142\vrb 2.55065\vrb 0.030\vrc
\+ \vra 3\vrb 0\vrb 4.23356\vrb 4.99255\vrb 3.05224\vrb 3.05118\vrb 0.035\vrc
\+ \vra 4\vrb 0\vrb 5.74962\vrb 6.80514\vrb 3.45341\vrb 3.45213\vrb 0.037\vrc
\+ \vra 5\vrb 0\vrb 7.26708\vrb 8.61823\vrb 3.79482\vrb 3.79336\vrb 0.039\vrc
\+ \vra 1\vrb 1\vrb 2.21348\vrb 2.37192\vrb 2.30073\vrb 2.30050\vrb 0.010\vrc
\+ \vra 2\vrb 1\vrb 3.68538\vrb 4.15501\vrb 2.85486\vrb 2.85434\vrb 0.018\vrc
\+ \vra 3\vrb 1\vrb 5.17774\vrb 5.95300\vrb 3.28659\vrb 3.28583\vrb 0.035\vrc
\+ \vra 4\vrb 1\vrb 6.67936\vrb 7.75701\vrb 3.64835\vrb 3.64739\vrb 0.026\vrc
\+ \vra 5\vrb 1\vrb 8.18607\vrb 9.56408\vrb 3.96382\vrb 3.96268\vrb 0.029\vrc
\+ \vra 1\vrb 2\vrb 3.21149\vrb 3.37018\vrb 2.65775\vrb 2.65756\vrb 0.007\vrc
\+ \vra 2\vrb 2\vrb 4.66860\vrb 5.14135\vrb 3.12077\vrb 3.12033\vrb 0.014\vrc
\+ \vra 3\vrb 2\vrb 6.14672\vrb 6.92911\vrb 3.50309\vrb 3.50245\vrb 0.018\vrc
\+ \vra 4\vrb 2\vrb 7.63639\vrb 8.72515\vrb 3.83336\vrb 3.83254\vrb 0.021\vrc
\+ \vra 5\vrb 2\vrb 9.13319\vrb 10.52596\vrb 4.12678\vrb 4.12581\vrb 0.024\vrc
\+ \vra 1\vrb 3\vrb 4.21044\vrb 4.36923\vrb 2.95461\vrb 2.95445\vrb 0.005\vrc
\+ \vra 2\vrb 3\vrb 5.65879\vrb 6.13298\vrb 3.35798\vrb 3.35759\vrb 0.012\vrc
\+ \vra 3\vrb 3\vrb 7.12686\vrb 7.91304\vrb 3.70327\vrb 3.70270\vrb 0.015\vrc
\+ \vra 4\vrb 3\vrb 8.60714\vrb 9.70236\vrb 4.00810\vrb 4.00737\vrb 0.018\vrc
\+ \vra 5\vrb 3\vrb 10.09555\vrb 11.49748\vrb 4.28283\vrb 4.28196\vrb 0.020\vrc
\+ \vra 1\vrb 4\vrb 5.20980\vrb 5.36863\vrb 3.21247\vrb 3.21233\vrb 0.004\vrc
\+ \vra 2\vrb 4\vrb 6.65235\vrb 7.12732\vrb 3.57310\vrb 3.57275\vrb 0.010\vrc
\+ \vra 3\vrb 4\vrb 8.11305\vrb 8.90148\vrb 3.88950\vrb 3.88898\vrb 0.013\vrc
\+ \vra 4\vrb 4\vrb 9.58587\vrb 10.68521\vrb 4.17335\vrb 4.17268\vrb 0.016\vrc
\+ \vra 5\vrb 4\vrb 11.06163\vrb 12.47532\vrb 4.43164\vrb 4.43131\vrb 0.008\vrc
}$$
\baselineskip 18 true pt 

\np
\hbox{\vbox{\psfig{figure=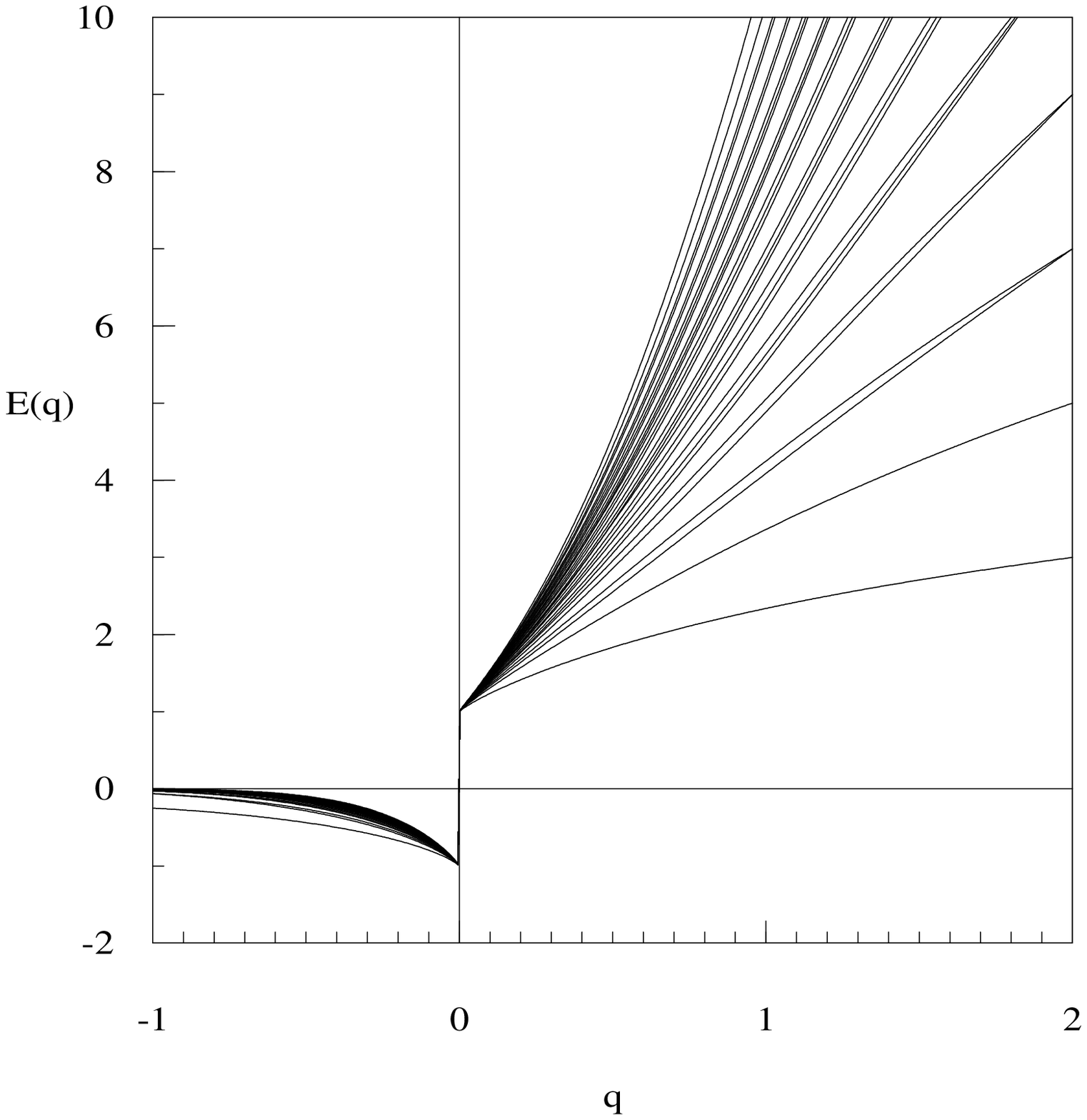,height=6in,width=5in,silent=}}}

\title{Figure 1.}
\nl The first 30 eigenvalues $E_{n\ell}(q),$ $1 \leq n \leq 5,$ $0 \leq \ell \leq 5,$ corresponding to the power potential $V(r) = \sgn(q) r^{q}.$  For $q > 0,$ the eigenvalues increase with $q$  from $1$ to $E_{n\ell}(2) = 4n + 2\ell + 1;$~  for $q < 0,$ they decrease (as $q$ increases) from $E_{n\ell}(-1) = -[2(n+\ell)]^{-2}.$ to $-1.$  Both sets of curves increase with $n$ and $\ell.$
\np
\hbox{\vbox{\psfig{figure=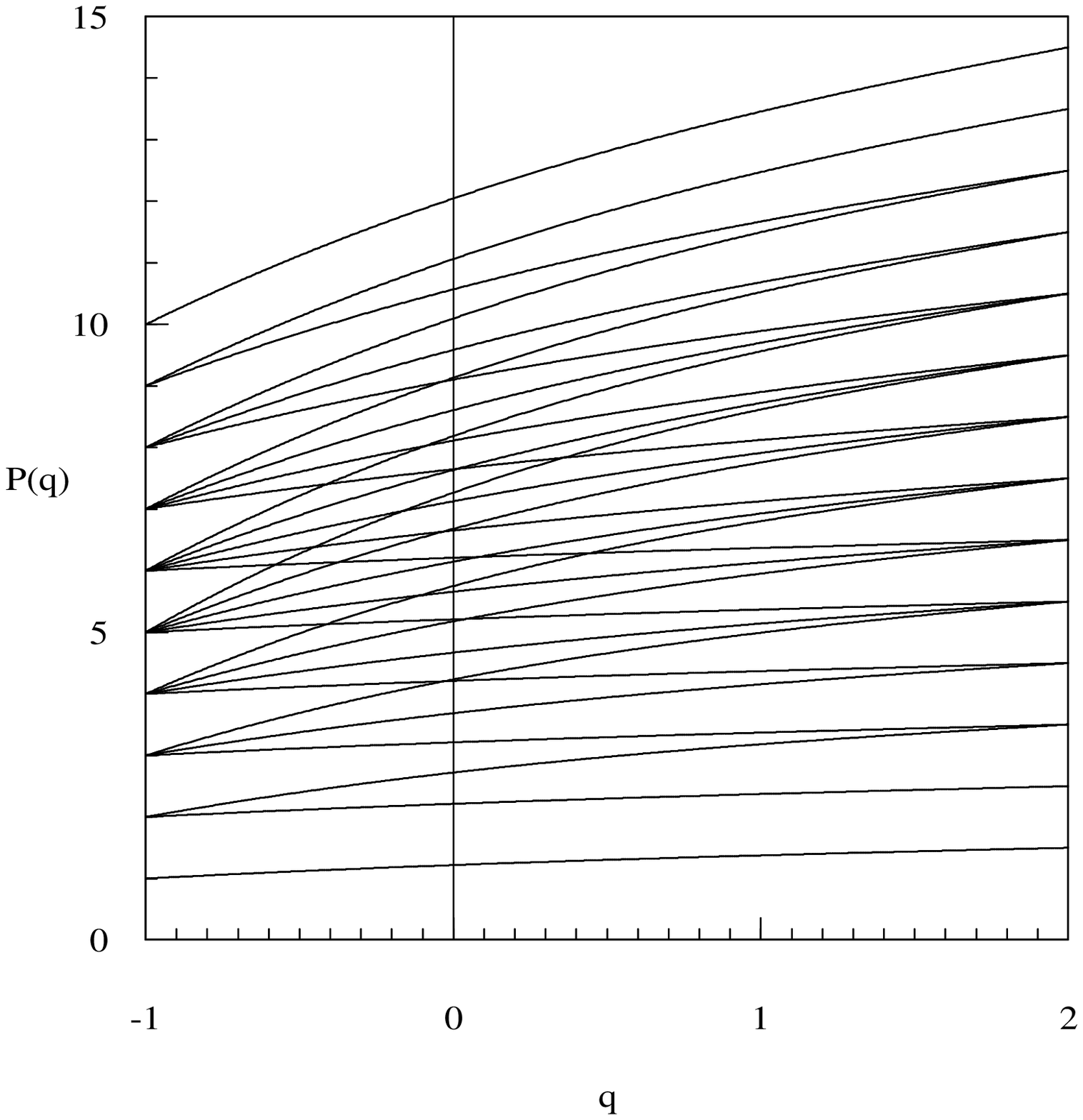,height=6in,width=5in,silent=}}}

\title{Figure 2.}
\nl In the $P$-representation, the same set of 30 eigenvalues shown in Fig.(1) now lie on monotone smooth curves.  The log-power theorem states that the $P$ values for the log potential are precisely $P_{n\ell}(0).$ As $q$ increases from $-1$ to $2$, the degeneracy of the Coulomb problem $P_{n\ell}(-1) = n + \ell$ evolves into the degeneracy of the harmonic oscillator $P_{n\ell}(2) = 2n + \ell - \half.$
\np
\hbox{\vbox{\psfig{figure=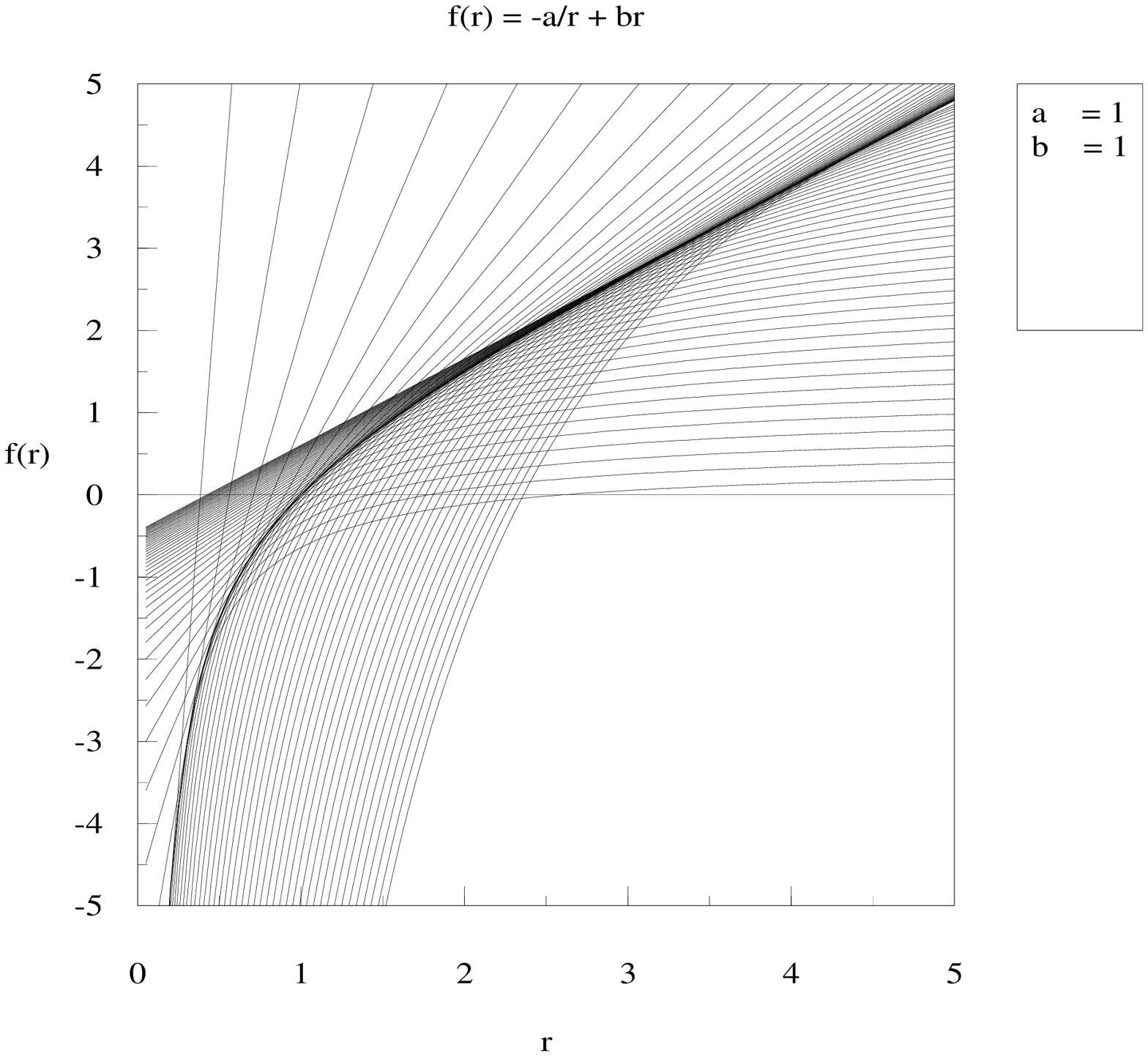,height=6in,width=5in,silent=}}}

\title{Figure 3.}
\nl The linear plus Coulomb potential shape $f(r) = -1/r + r$ represented as the envelope curve of two distinct families of potentials of the form $\alpha h(r) + \beta.$  In the upper family $h(r) = r$ is a linear potential; in the lower family $h(r) = -1/r$ is a Coulomb potential.
\np
\hbox{\vbox{\psfig{figure=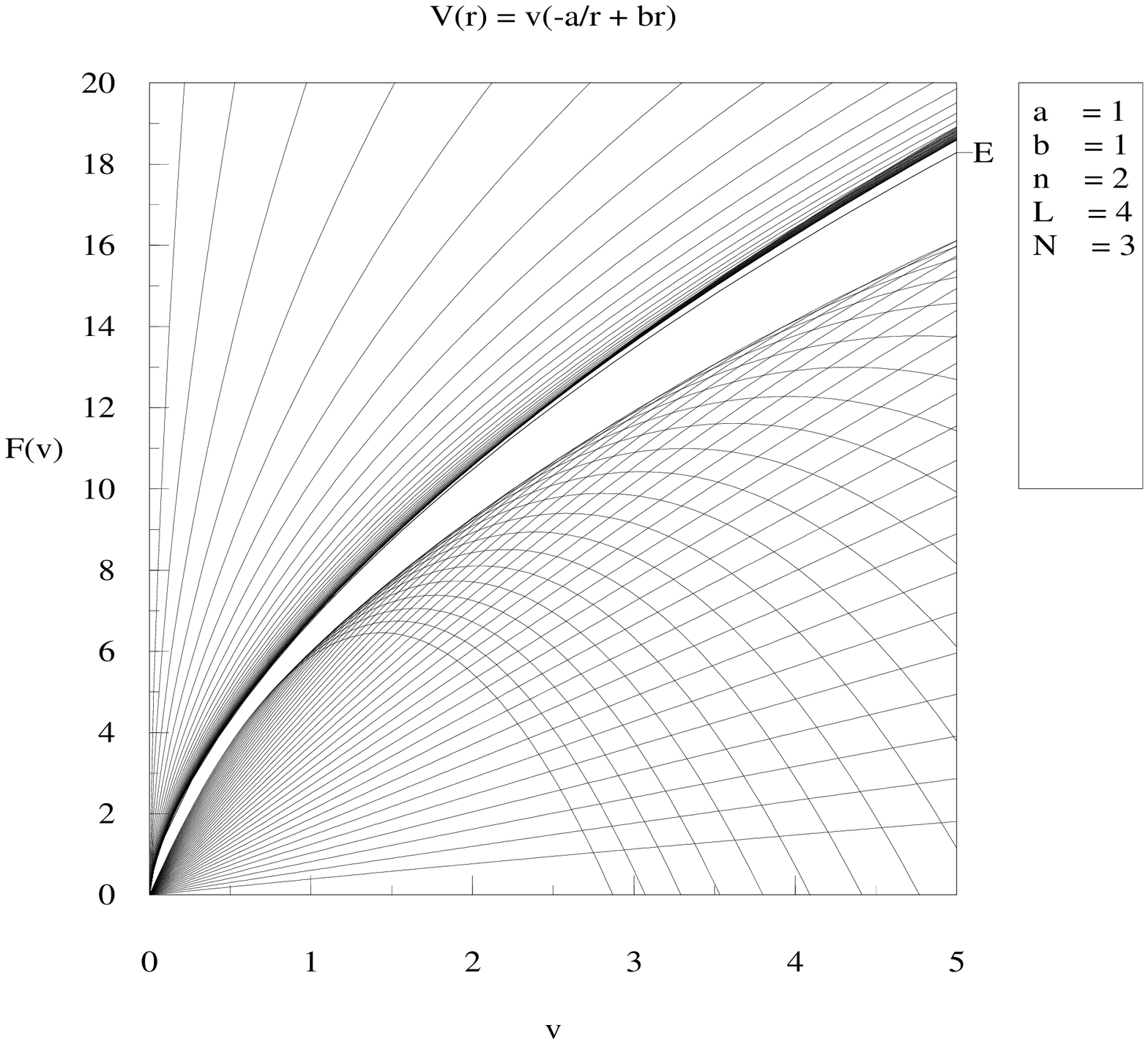,height=6in,width=5in,silent=}}}

\title{Figure 4.}
\nl The spectral approximation corresponding to Fig.(3).  Each `tangential' potential $f^{(t)}(r) = \alpha h(r) + \beta$ generates a corresponding tangential energy curve $F^{(t)}(v) = H(\alpha v) + \beta v.$  The envelopes of these spectral families generate upper and lower bounds to the exact curve $E = F(v),$ shown here for the case $(n, \ell) = (2, 4).$
\np
\hbox{\vbox{\psfig{figure=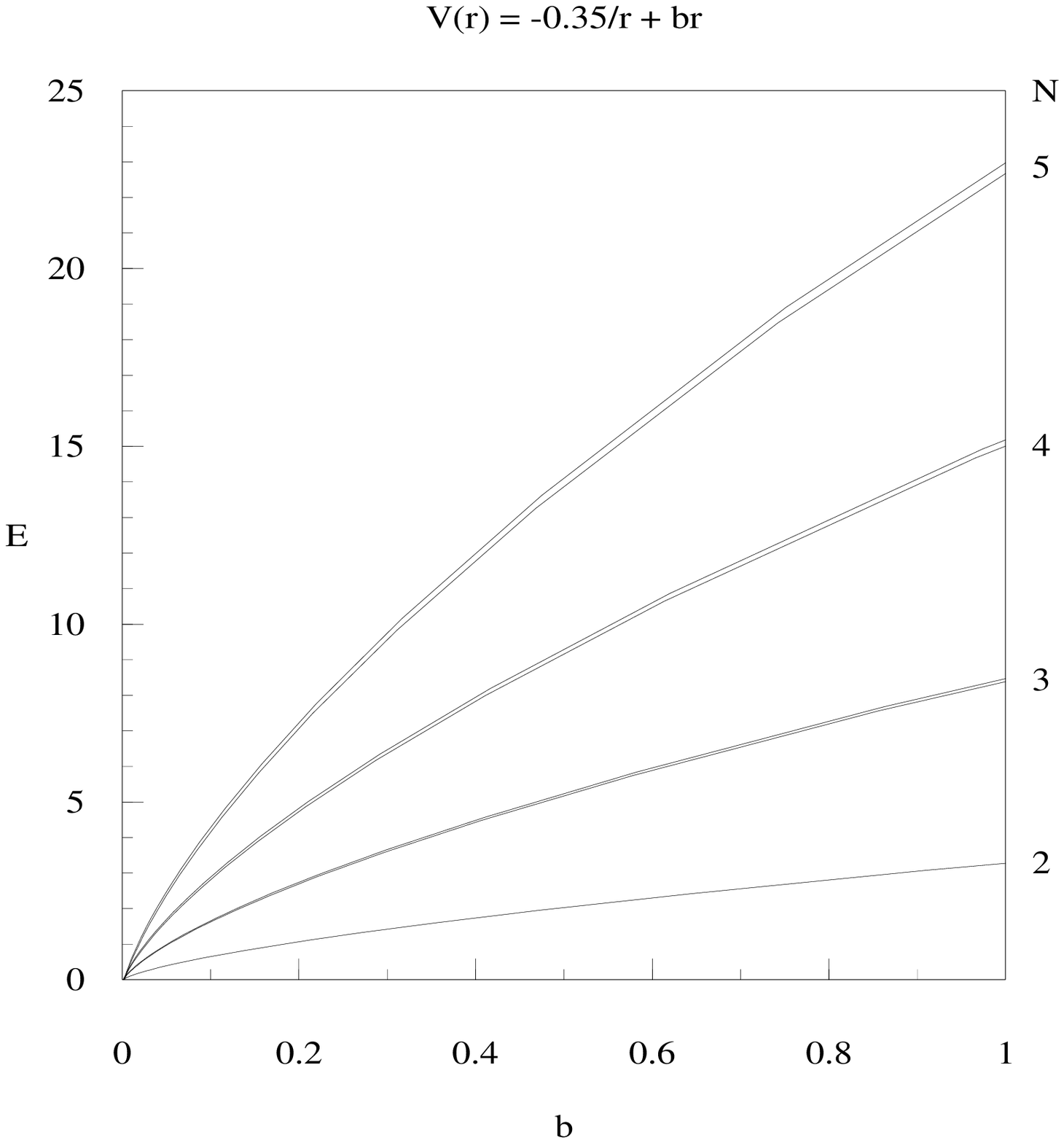,height=6in,width=5in,silent=}}}

\title{Figure 5.}
\nl In a naive model with linear plus Coulomb pair potentials, the graphs show upper and lower bounds to the energies of the lowest spatially many-body symmetric \hi{N}{quark} states.  These states are only physically accessible for $N\leq 3,$ but the bounds are valid for all $N\geq 2.$ In the case of $N = 2$ there is only one curve because the solution is `exact'.

\np
\hbox{\vbox{\psfig{figure=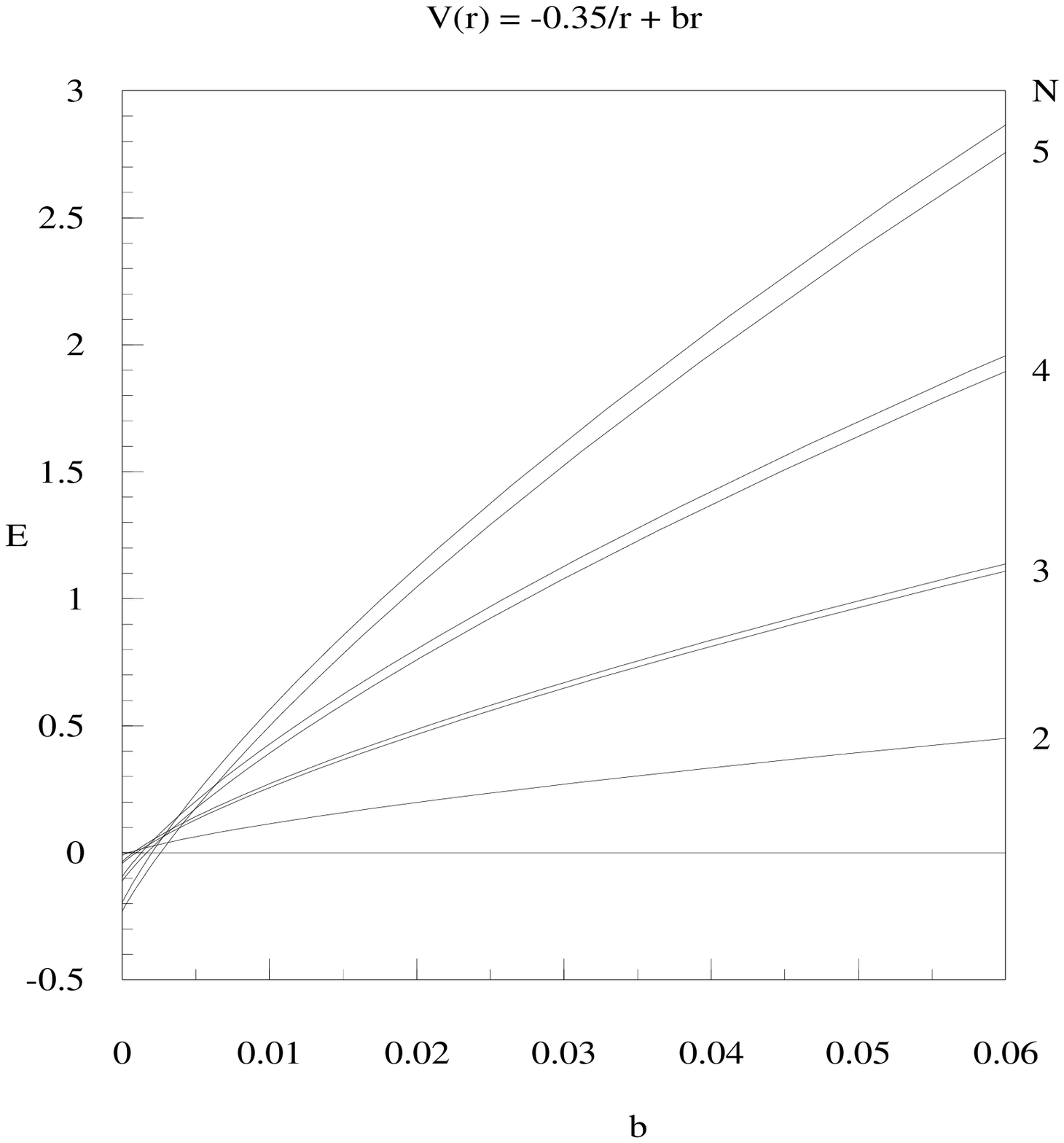,height=6in,width=5in,silent=}}}

\title{Figure 6.}
\nl The many-body graphs are the same as in Fig.(5) but with a smaller scale showing the intersection of the curves.  Graphs such as this may help with the elementary consistency question: can we devise a potential that yields the known \hi{2}{} and \hi{3}{quark} masses?

\end